\documentclass[12pt]{article}
\usepackage{setspace}
\usepackage[utf8]{inputenc}
\usepackage{geometry}
\usepackage{graphicx}
\usepackage{booktabs}
\usepackage{array}
\usepackage{paralist}
\usepackage{verbatim}
\usepackage{subfig}
\usepackage{amsmath}
\usepackage{float}
\usepackage{pgfplots} 
\usepackage{pgfplotstable}
\usepackage{filecontents}
\usepackage{multirow}
\usepackage[title]{appendix}
\usepackage[cbgreek]{textgreek}
\usetikzlibrary{matrix}
\usepgfplotslibrary{groupplots}
\pgfplotsset{compat=1.14}
\definecolor{bblue}{HTML}{4F81BD}
\definecolor{rred}{HTML}{C0504D}
\definecolor{ggreen}{HTML}{9BBB59}
\definecolor{ppurple}{HTML}{9F4C7C}
\definecolor{mypink}{cmyk}{0.0, 0.7808, 0.4429, 0.1412}
\definecolor{mygray}{gray}{0.3}
\definecolor{gray1}{gray}{0.5}
\definecolor{gray2}{gray}{0.7}
\definecolor{gray3}{gray}{0.9}
\definecolor{orange}{cmyk}{0.0, 0.3, 0.6, 0.0}
\definecolor{orange3}{cmyk}{0.0, 0.1, 0.2, 0.0}
\definecolor{orange2}{cmyk}{0.0, 0.3, 0.6, 0.0}
\definecolor{orange1}{cmyk}{0.0, 0.5, 1.0, 0.0}
\definecolor{blue1}{rgb}{0.4, 0.4, 0.6}
\definecolor{blue2}{rgb}{0.55, 0.55, 0.8}
\definecolor{blue3}{rgb}{0.7, 0.7, 1.0}
\definecolor{green1}{rgb}{0.0, 0.6, 0.0}
\definecolor{green2}{rgb}{0.0, 0.8, 0.0}
\definecolor{green3}{rgb}{0.0, 1.0, 0.0}
\definecolor{yellow1}{rgb}{0.6, 0.6, 0.0}
\definecolor{yellow2}{rgb}{0.8, 0.8, 0.0}
\definecolor{yellow3}{rgb}{1.0, 1.0, 0.0}
\definecolor{red1}{rgb}{0.6, 0.2, 0.2}
\definecolor{red2}{rgb}{0.8, 0.3, 0.3}
\definecolor{red3}{rgb}{1.0, 0.4, 0.4}

\usepackage{fancyhdr}
\usepackage{hyperref}

\usepackage{sectsty}
\allsectionsfont{\sffamily\mdseries\upshape}
\usepackage[nottoc,notlof,notlot]{tocbibind}
\usepackage[titles,subfigure]{tocloft}

\usepackage{algorithm}
\usepackage{algorithmic}
\usepackage{amssymb}

\title{Reverse Quantum Annealing Approach \\to Portfolio Optimization Problems}
\author{Davide Venturelli$^{1}$ \and Alexei Kondratyev$^2$}
\date{%
    \small{$^1$\it{USRA Research Institute for Advanced Computer Science,
    Mountain View, CA 94035, USA
    }}\\
    \small{$^2$\it{Data Analytics Group, Financial Markets, Standard Chartered Bank, 
    London EC2V 5DD, UK}}\\[2ex]%
}

\begin{document}
\singlespacing

\maketitle

\vskip2cm

\abstract{
We investigate a hybrid quantum-classical solution method to the mean-variance portfolio optimization problems. 
Starting from real financial data statistics and following the principles of the Modern Portfolio Theory, we 
generate parametrized samples of portfolio optimization problems that can be related to quadratic binary 
optimization forms programmable in the analog D-Wave Quantum Annealer 2000Q\textsuperscript{TM}. The instances 
are also solvable by an industry-established Genetic Algorithm approach, which we use as a classical benchmark. 
We investigate several options to run the quantum computation optimally, ultimately discovering that the best 
results in terms of expected time-to-solution as a function of number of variables for the hardest instances 
set are obtained by seeding the quantum annealer with a solution candidate found by a greedy local search 
and then performing a reverse annealing protocol. The optimized reverse annealing protocol is found to be more than 100 times faster than the
corresponding forward quantum annealing on average.
}

\vskip3cm

\noindent
\begin{tabular}{l l}
	\bf{Keywords:}& Portfolio Optimization, Quadratic Unconstrained Binary \\
				  & Optimization, Quantum Annealing, Genetic Algorithm,     \\
				  & Reverse Quantum Annealing                              \\
				  &                                                        \\
	\bf{JEL classification:}	& C61, C63, G11                            \\
\end{tabular}

\newpage

\section{Introduction} \label{sec:intro}
One reason behind the current massive investment in quantum computing is that this computational 
paradigm has access to resources and techniques that are able to circumvent several bottlenecks of 
state-of-the-art digital algorithms.  Much research is today devoted to understanding how to exploit
this power to deliver high quality solutions to discrete optimization problems at a fraction of the 
time required using the best classical methods on high-end hardware.

Quantum annealers are special-purpose machines inspired by the adiabatic quantum computing paradigm.
These machines are manufactured by D-Wave Systems and appeared on the market in 2011, and while being
limited in programmability with respect to other experimental devices under testing by other companies, 
are still the only available quantum devices that feature a sufficient amount of quantum memory (qubits) 
to be applied to non-trivial problems at the time of writing. For this reason they are subject to extensive 
empirical investigation by several groups around the world, not only for scientific and research purposes
\cite{ronnow2014defining, denchev2016computational}, but also for performance evaluation on structured real 
world optimization challenges \cite{venturelli2015quantum, stollenwerk2017quantum, mott2017solving}. It is expected that as quantum computing devices mature, they'd become available to enterprises for use through dedicated cloud services~\cite{patentQCware}.

Optimization problems comprise a large class of hard to solve financial problems, not to mention the fact 
that many supervised and reinforcement learning tools used in finance are trained via solving optimization 
problems (minimization of a cost function, maximization of reward). Several proposed applications of the 
D-Wave machine to portfolio optimization \cite{rosenberg2016solving, marzec2016portfolio}, dealt with 
portfolios that were too small 
to evaluate the scaling of the chosen approach with the problem size. In this paper we go beyond these early 
approaches and provide an analysis on sufficient data points to infer a scaling and measure a limited 
speedup with respect a state-of-the-art numerical method based on genetic algorithms. Moreover, we 
extensively tune the D-Wave machine's runs using the hybrid method of Reverse Annealing which is available 
only on the most advanced models to date, and whose usage has never been published before on structured 
optimization problems.

\paragraph{Portfolio optimization}
The optimal portfolio construction problem is one of the most extensively studied problems in 
quantitative finance. The Modern Portfolio Theory (MPT) as formulated by Markowitz \cite{markowitz1952portfolio} 
has laid the foundation for highly influential mean-variance portfolio optimisation approach. According 
to the MPT, a typical portfolio optimisation problem can be formulated as follows. Let $N$ be the number
of assets, $\mu_i$ be the expected return of asset $i$, $\sigma_{ij}$ be the covariance between the returns
for assets $i$ and $j$, and $R$ be the target portfolio return. Then the decision variables are the weights 
$w_i$, i.e. the investment associated with asset $i$ ($w_i \in \mathbb{R}$). The standard Markowitz 
mean-variance approach consists in the constrained, quadratic optimisation problem:
\begin{eqnarray} \label{eq:MarObjFun}
\min \sum_{i=1}^{N} \sum_{j=1}^{N} w_i w_j \sigma_{ij}\;\;;&\sum_{i=1}^{N} 
w_i = 1\;\;;&\sum_{i=1}^{N} w_i \mu_i = R\;.
\end{eqnarray}

Quadratic problems of the form (\ref{eq:MarObjFun}) are efficiently solvable 
by standard computational means (e.g., quadratic programming with linear
constraints \cite{fernando2000practical}) if the covariance matrix is positive definite. 
Interestingly, the problem can also be cast into an unconstrained quadratic
optimization problem (QUBO) which is a suitable formulation for quantum
annealers. This observation spurred in the last few years a couple of
proof-of-principle papers that were performing runs on the first generation
D-Wave machines solving discrete portfolio optimization problems
\cite{rosenberg2016solving, marzec2016portfolio}. The discrete portfolio optimization problem is much
harder to solve than continuous mean-variance portfolio optimization problem 
due to strong non-linearity, and is known to be NP-complete \cite{kellerer2000selecting}.

In this paper we propose an original approach to portfolio optimization that extends the
mean-variance model to a general dependence structure and allows portfolio
managers to express discretionary views on relative attractiveness of different assets and 
their combinations. The manuscript is structured as follows: in Section \ref{sec:problem} the
formulation of the portfolio optimization problem is constructively described starting from real
market data. Section \ref{sec:annealer} describes the hybrid quantum annealing solver and its
setup. Section \ref{sec:main} is devoted to the experimental results, including results 
obtained with the classical Genetic Algorithm (GA) benchmark (following \cite{Kondratyev2017risk}). 
In Section \ref{sec:conclusion} we conclude with a discussion and  considerations for future work.\\

\section{Portfolio Optimization Beyond Markowitz} \label{sec:problem}

The problem we are trying to solve here is construction of the optimal portfolio from the 
assets with known characteristics such as asset returns, volatilities and pairwise correlations.
A typical portfolio optimization problem consists of selecting $M$ assets from
the universe of $N$ investible assets. These $M$ assets should ideally be the
best possible choice according to some criteria. The scenario we target is a
Fund of Funds portfolio manager who is facing a task of selecting the best funds
that follow particular trading strategies (e.g., EM Macro Funds) in order to
maximize the risk-adjusted returns according to some model.

A realistic case occurs when the assets are selected with equal preference weights (a classical  
example of exposure to equally weighted assets is the CSO\footnote{Collateralized Synthetic Obligation 
(CSO) is a type of Collateralized Debt Obligation (CDO) where credit exposure to the reference names 
is provided in synthetic form via single name Credit Default Swaps (CDS). A typical CSO references 
between 100 and 125 equally weighted names.} portfolio). Should we want to generalize the portfolio 
with larger allocation to a given asset, we could allow for multiples of the reference weight by 
cloning an asset and treating it as a new one.

The task of encoding the relationship among the choices of $M$ funds ({\it without replacement}) 
from the universe of $N$ funds can then be formulated as a quadratic form:
\begin{equation} \label{eq:QUBO}
O({\bf q}) = \sum_{i=1}^{N} a_i q_i + \sum_{i=1}^{N} \sum_{j=i+1}^{N} b_{ij} q_i q_j \; ,
\end{equation}

\noindent
where $q_i = 1$ means that asset $i$ is selected and $q_i = 0$ means 
that asset $i$ is not selected. Coefficients $a_i$ reflect asset attractiveness on a standalone
basis and can be derived from the individual asset's expected risk-adjusted returns. Coefficients
$b_{ij}$ reflect pairwise diversification penalties (positive values) and rewards (negative 
values). These coefficients can be derived from the pairwise correlations.

The minimization of the QUBO objective function given by expression
(\ref{eq:QUBO}) should optimize the risk-adjusted returns by the use of the
metrics of Sharpe ratio \cite{sharpe1966mutual} for the $a_i$ parameters (measuring 
expected excess asset return in the asset volatility units) and correlation
between assets (as a measure of diversification) for the $b_{ij}$ coefficients.

The Sharpe ratio is calculated as $(r-r_0)/\sigma$ where $r$ is the asset's
expected annualized return, $r_0$ is the applicable risk-free interest rate 
and $\sigma$ is asset's volatility (annualized standard deviation of the asset's
log-returns). The higher fund's Sharpe ratio, the better fund's returns have
been relative to the risk it has taken on. Volatility can be estimated as
historical annualized standard deviation of the Net Asset Value (NAV) per share
log-returns. Expected return can be either estimated as historical return on
fund investment or derived independently by the analyst/portfolio manager taking
into account different considerations about the future fund performance. The
asset correlation matrix $\rho_{ij}$ is constructed from the time series of
log-returns \cite{hull2016options}.

\paragraph{QUBO instances}
Instead of using the raw real numbers obtained from financial data for the QUBO
coefficients, we opt to coarse-grain the individual funds Sharpe ratios and
their mutual correlation values down to integer values by grouping intervals 
in buckets according to Table \ref{tab:tableBuckets}. 
By using bucketed values we define a scorecard, which is loosely based on
the past fund performances but can be easily adjusted by portfolio manager
according to his/her personal views and any new information not yet reflected in
the funds reports. Indeed, in real world scenarios, optimization needs to be
performed with respect to the discretionary views about the future held by
portfolio manager, not necessarily with respect to automatically fetched market
data. Moreover, if we assume that funds report their NAV per share on a monthly
basis and we have comparable funds data for the previous year, then we only have
12 NAV observations in our time series from which we want to construct a
correlation matrix for $N$ assets ($N \gg 12$). It is likely that the resulting
correlation matrix will not be positive definite due to the number of
observations being much smaller than the number of assets, ruling out the
traditional solvers of mean-variance portfolio optimization problems which
require a positive definite correlation matrix.\\ 

\begin{table}[!htb]
\footnotesize
\begin{minipage}{.5\linewidth}
\centering
\begin{tabular}{@{}*3c@{}}
  \toprule
  \multicolumn{1}{c} {\bf SR bucket (asset $i$)}  & \multicolumn{1}{c} {\bf Coefficient $a_i$} \\
                     Equally spaced buckets,      & Sample Mapping                             \\
                     from worst to best           & Scheme                                     \\
  \midrule
   1st   &        15 \\
   2nd   &        12 \\
   3rd   &         9 \\
   4th   &         6 \\
   5th   &         3 \\
   6th   &         0 \\
   7th   &      $-$3 \\
   8th   &      $-$6 \\
   9th   &      $-$9 \\
  10th   &     $-$12 \\
  11th   &     $-$15 \\
  \bottomrule
\end{tabular}
\end{minipage}%
\begin{minipage}{.5\linewidth}
\centering
\begin{tabular}{@{}*3c@{}}
  \toprule
  \multicolumn{1}{c} {\bf Correlation bucket}  & \multicolumn{1}{c} {\bf Coefficient $b_{ij}$} \\
                                               & Sample Mapping                                \\
                                               & Scheme                                        \\
  \midrule
  $-1.00 \leq \rho_{ij}    < -0.25$   &  $-$5   \\
  $-0.25 \leq \rho_{ij}    < -0.15$   &  $-$3   \\
  $-0.15 \leq \rho_{ij}    < -0.05$   &  $-$1   \\
  $-0.05 \leq \rho_{ij}    <  0.05$   &     0   \\
  $ 0.05 \leq \rho_{ij}    <  0.15$   &     1   \\
  $ 0.15 \leq \rho_{ij}    <  0.25$   &     3   \\
  $ 0.25 \leq \rho_{ij} \leq  1.00$   &     5   \\
  \bottomrule
\end{tabular}
\end{minipage} 
\caption{\small{Specification of the sample QUBO coefficients from NAV time series market data}}
\label{tab:tableBuckets}
\end{table}

The instance set used for our case study is obtained by simulating asset values with the help of 
correlated Geometric Brownian Motion (GBM) processes with uniform constant asset correlation $\rho$, drift 
$\mu$ and log-normal asset volatility $\sigma$. The specific values of these parameters were derived 
from a wide range of fund industry researches (see, e.g., \cite{darolles2010conditionally} for the Sharpe ratio 
distributions) and, therefore, can be viewed as representative for the industry. The simulation time 
horizon was chosen to be 1 year and the time step was set at 1 month $-$ this setup corresponds to 
the situation where a Fund of Funds portfolio manager is dealing with the short time series of 
comparable monthly fund performance reports. Detailed description of the simulation scheme can be 
found in Appendix \ref{sec:gbm}. 
Every simulated (or, "realized") portfolio scenario consists of 12 monthly returns for each asset. 
From these returns we calculated total realized return and realized volatility for each asset 
(which, obviously, differ from the their expected values $\mu$ and $\sigma$) and for the portfolio 
as a whole. We also calculated realized pairwise correlations between all assets
according to the input uniform correlation $\rho$. 
Finally, we calculated individual assets and portfolio Sharpe ratios. 
For instance, with $\rho = 0.1$, $\mu = 0.075$, $\sigma = 0.15$ and the constant risk-free interest 
rate set at $r_0 = 0.015$, the expected Sharpe ratio for each asset in a portfolio is 0.4. The expected 
Sharpe ratio for the portfolio of $N$ assets is significantly larger due to the diversification and low 
correlation between the assets, e.g., for a 48-asset portfolio we would expect Sharpe ratio
values from 0.5 (25th percentile) to 2.1 (75th percentile) with a mean around 1.4.

\paragraph{Constraint on the number of selected assets} In QUBOs, the standard way to deal 
with the cardinality constraint (i.e., selection of only $M$ assets) would be to add a term 
$O_{\textrm{penalty}}({\bf q})$ to the objective function given by expression (\ref{eq:QUBO}) 
such that the unsatisfying selections would be penalized by a large value $P \gg 1$, which would 
force the global minimum to be such that $\sum_{i=1}^N q_i=M$:
\begin{equation} \label{eq:Penalty}
O_{\textrm{penalty}}({\bf q}) = P \left( M - \sum_{i=1}^{N} q_i \right)^{2} \; .
\end{equation}

Unfortunately, the introduction of such a large energy scale $P$ is typically associated with 
precision issues connected to the analog nature of the machine and the fact that there is a 
physical maximum to the energy that can be controllably programmed on local elements of the 
D-Wave chip~\cite{johnson2011quantum}. However, several hybrid quantum-classical strategies can be put in place
to overcome the limitation. For instance, we observe that shifting artificially the Sharpe ratio values 
by a constant amount $\pm \Delta$ (and adding buckets according to the prescription chosen, 
e.g., Table \ref{tab:tableBuckets}) will essentially amount to forcing the ground state solution 
of the unconstrained problem to have more or less desired number of assets selected. 
Hence, while not solving the same problem, we could imagine a solver 
of a similarly constrained problem that will enclose the quantum annealing runs in a classical loop 
which checks for the number of selected assets $m(\Delta)$ in the best found solution with $\Delta=0$, 
then increases or decreases the individual desirability of the assets according to whether 
$m$ is larger or smaller than $M$ and runs again until $m(\Delta)=M$ for $\Delta=\Delta^\star$. 
This sort of hybridization scheme is not uncommon for quantum-assisted solvers~\cite{tran2016hybrid,venturelli2015quantum} and the number 
of expected rounds of runs should scale as $\propto \log_2(\Delta^\star)$ as per a binary search 
introducing a prefactor over the time-to-solution complexity which should stay manageable. 
Other hybrid approaches could also be put forward to deal with the constraint such as fixing 
some asset selections in pre-processing via sample persistence~\cite{karimi2017boosting}.

Per above arguments, in our benchmark case study in this work, we will focus on running 
unconstrained problems, setting $\Delta=0$. By design, the instance ensambles are such that the 
average number of assets in the ground state should be around $N/2$ 
(see Table~\ref{tab:GreedySearchAndParams}) corresponding to the 
largest combinatorial space by exhaustive enumeration of the possible solutions, so our 
tests will arguably be targeting among the most challenging instances of our parametrized model.\\

\section{Quantum Annealing Hybrid Solver} \label{sec:annealer}

A QUBO problem can be easily translated into a corresponding Ising problem
of $N$ variables $s_i$, $i = 1,\ldots, N$ with $s_i \in \{+1, -1\}$ given by the following expression:
\begin{equation} \label{eq:IsingObjFun}
O_{\textrm{Ising}}({\bf s}) = \sum_{i=1}^{N} h_i s_i + \sum_{i=1}^{N} 
\sum_{j=i+1}^{N} J_{ij} s_i s_j \; .
\end{equation}

Ising and QUBO models are related through the transformation $s_i = 2 q_i - 1$, hence the relationship with Eq.(\ref{eq:QUBO}) is $J_{ij}=\frac{1}{4}b_{ij}$ and $h_i=(\frac{1}{2}a_i+\sum_{j}b_{ij})$, disregarding an 
unimportant constant offset.

\paragraph{Chimera-embedded Ising representation}
In the standard programming practices of the analog D-Wave Quantum Annealer 2000Q\textsuperscript{TM} 
(DW2000Q) each spin-variable $s_i$ should be ideally assigned to a specific chip element, a superconducting 
flux qubit, modeled by a quantum two level system that could represent the quantum Hamiltonian $\mathcal{H}_{\textrm{local}}=\sum_i  h_i \sigma_i^z$, with $\sigma_i^z$ being the usual Pauli matrix 
representation for an Ising quantum spin-$\frac{1}{2}$. Each qubit supports the programming of the 
$h_i$ terms. Instead, $J_{ij}$ parameters can then be implemented energetically through inductive elements, meant to
represent $\mathcal{H}_{\textrm{couplers}}=\sum_{ij}J_{ij}\sigma_i^z\sigma_j^z$, if and only if the required 
circuitry exists between qubits $i$ and $j$. Couplers cannot be manufactured connecting qubits too far apart in the spatial layout 
of the processor due to engineering considerations. In other words $J_{ij}=0$ unless $(i,j)\in \chi_{16}$, 
where $\chi_{16}=(V,E)$ is the \emph{Chimera} graph of DW2000Q, featuring 16x16 unit cells of 8 qubits each, 
in the ideal case, for a total of 2048 qubits
\footnote{note that the graph is not ideal, there is a set of 
17 qubits that have not been calibrated successfully and are 
unoperable (See Fig.~\ref{fig:chimera} 
in Appendix \ref{appendixEmbedding})}.

To get around this restriction, we employ the minor-embedding compilation technique for fully connected graphs~\cite{boothby2016fast,venturelli2015spinglass}. 
By means of this procedure, we obtain another Ising form, where qubits are arranged in ordered 1D chains 
interlaced on the Chimera graph:
\begin{eqnarray} 
\mathcal{H}_{\chi\textrm{-Ising}}&=&-\sum_{i=1}^N |J_F| \left[ \sum_{c=1}^{N_c-1} \bf{\sigma_{ic}^z} \sigma_{i(c+1)}^z\right]\label{eq:IsingEmbeddedObjFunJF}\\
&+&\sum_{i=1}^{N} \frac{h_i}{N_c} \left[\sum_{c=1}^{N_c} \bf{\sigma_{ic}^z}\right]
+ \sum_{i,j=1}^{N} J_{ij} \sum_{c_i,c_j=1}^{N_c} \delta_{ij}^\chi(c_i,c_j) \bf{\sigma_{ic_i}^z}\bf{\sigma_{jc_j}^z} \; .
\label{eq:IsingEmbeddedObjFun}
\end{eqnarray}

In Eq.~(\ref{eq:IsingEmbeddedObjFunJF}) we explicitly isolate the encoding of the \emph{logical} quantum
variable: the classical binary variable $s_i$ is associated with $N_c$=($N/4+1$) Ising spins $\sigma_{ic}^z$, ferromagnetically coupled directly by strength $J_F$, forming an ordered 1D chain subgraph of $\chi_{16}$. 
The value of $J_F$ should be strong enough to correlate the value of the magnetization of each individual 
spin if measured in the computational basis ($\langle\bf{\sigma_{ic}^z}\rangle=\langle\bf{\sigma_{ic^\prime}^z}\rangle$).
In the term (\ref{eq:IsingEmbeddedObjFun}), we encode the objective function (\ref{eq:IsingObjFun}) through 
our extended set of variables: the local field $h_i$ is evenly distributed across all qubits belonging to the 
logical chain $i$, and each couplers $J_{ij}$ is active only between one specific pair of qubits
$\bf{\sigma_{ic_i^\star}^z}$, $\bf{\sigma_{jc_j^\star}^z}$ which is specified by the adjacency check function
$\delta^\chi_{ij}(c_i,c_j)$ which assumes unit value only if $(c_i=c_i^\star)$ and $(c_j=c_j^\star)$ 
and is zero otherwise\footnote{In the actual embedding employed, it might happen that some pairs of 
logical variables $i$,$j$ could have two pairs that can be coupled, instead of one. In that case, we 
activate both couplings at a strength $J_{ij}/2$ to preserve the classical value of the objective function.}.

\begin{figure}
\begin{center}
\includegraphics[width=0.9\columnwidth]{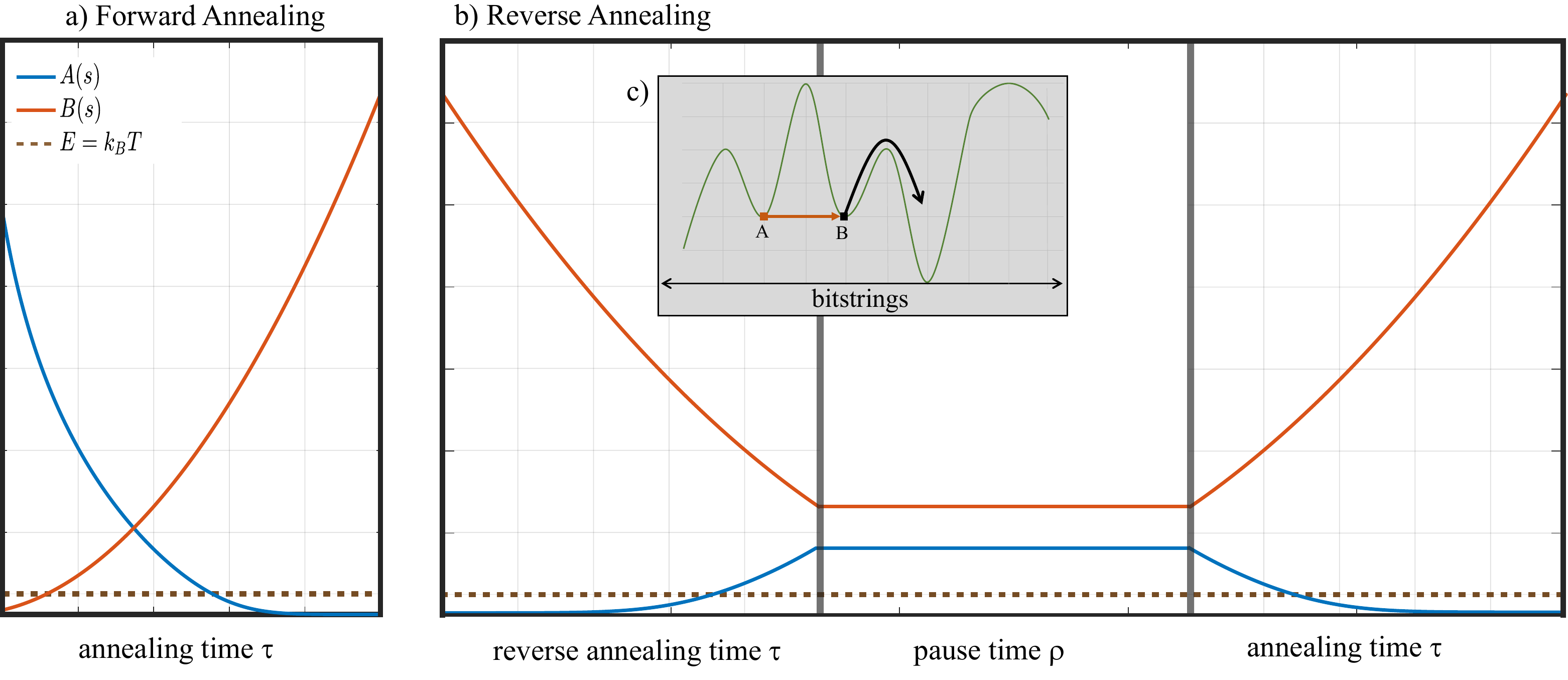}
\caption{\small{Quantum annealing schedules on DW2000Q used in our
investigations, with reference to Eq.(\ref{eq:QA}). 
a) Forward quantum annealing schedule. b) Reverse quantum annealing schedule
where the three steps of the protocol are all set to the same duration of
$\tau$=$\rho$. c) Pictorial view of what should be facilitated by reverse
quantum annealing: when the initial bitstring is a local minimum (A), the
reverse protocol should allow tunneling between valleys (A$\rightarrow$B) 
while the pause should facilitate gradient descent as well as allow for local
exploration through thermal hopping (escape from B towards the global minimum).}
}
\label{fig:schedule}
\end{center}
\end{figure}

\paragraph{Forward Quantum Annealing}
The quantum annealing protocol, inspired by the adiabatic principle of quantum mechanics, dictates to 
drive the system from an initial (easy to prepare) ground state of a quantum Hamiltonian $\mathcal{H}_{\textrm{initial}}$ to the unknown low energy subspace of states of the problem Hamiltonian
(\ref{eq:IsingEmbeddedObjFun}), ideally to the lowest energy states corresponding to the global minima 
of the objective function (\ref{eq:IsingObjFun}).
This \emph{forward} quantum annealing procedure on DW2000Q can be ideally described as 
attempting to drive the evolution of the following time-dependent Hamiltonian:
\begin{eqnarray}
\mathcal{H}_{\textrm{QA}}(t) &=& A[t] \sum_{i=1}^N\left(\sum_{c=1}^{N_c} \bf{\sigma_{ic}^x}\right) + B[t] \mathcal{H}_{\chi\textrm{-Ising}} \; ,
\label{eq:QA}
\end{eqnarray}

\noindent
where the term multiplying $A[t]$ is  a Hamiltonian describing 
an independent collection of local transverse fields for each spin of the system~\cite{johnson2011quantum}. 
Figure~\ref{fig:schedule}-a shows how $A[t]$ and $B[t]$ are varying over the scale of 
total \emph{annealing time} $\tau$ on DW2000Q in units of the typical energy scale of the nominal
temperature measured on the chip, 12 milliKelvin. At the end of the annealing run, $A[t]$=0 and
the system is projected on the computational basis by a measurement of each qubit magnetization.
The duration of the anneal, $\tau$, is a free parameter, hence it is often useful to define the 
fractional completion of the annealing schedule, $s=t/\tau$.

We note that the real time-dependent evolution of the spin-system programmed in the DW2000Q chip
is only loosely modeled correctly by the Schr\"odinger evolution of $\mathcal{H}_{\textrm{QA}}$, 
mostly because of the dominance of open-system dynamics effects~\cite{job2018test}, the existence 
of unknown static and dynamic misspecification on the programmed parameters $h_i$ and $J_{ij}$, 
and cross-talk couplings~\cite{king2014algorithm}. Recent research attempts to model accurately 
the forward annealing dynamics of DW2000Q and earlier models of D-Wave machines by means of ab-initio 
reasoning have been moderately successful on carefully crafted problems~\cite{boixo2016computational}. 
However, for the purpose of benchmarking the details of the evolution are not important: once specified
$\mathcal{H}_{\chi-\textrm{Ising}}$, a single machine run could essentially be treated as a black-box 
solver dependent on the parameters $J_{F}$, and $\tau$, returning a bitstring at every run.

\paragraph{Reverse Quantum Annealing} 
Figure \ref{fig:schedule}-b illustrates the quantum annealing protocol when the DW2000Q 
is set to operate as a reverse annealer. The system is initialized with $B[t]$=$\max{B}$ 
and $A[t]$=0, with spins set to a classical bitstring $\mathcal{S}$. The evolution
then undergoes an inverse schedule illustrated in Figure \ref{fig:schedule}-a 
up to a point where the Hamiltonian time-dependence is temporarily paused. With
reference to the Hamiltonian evolution in Eq.(\ref{eq:QA}), the transverse field
evolution that we program\footnote{Many options are possible, since the duration
of the three phases can be choosen arbitrarily within limited but wide ranges.}
for this protocol is the following three-phases function (identical equations
for $B^R[t]$, $B^p[t]$, $B^F[t]$ in terms of $B[t]$):
\begin{eqnarray}
A^R[t]=A[\tau+(s_p-1)t]&\;\;&\textrm{(Reverse Annealing)}\;\; t\in[0,\tau]\label{eq:reverseSchedule}\\
A^p[t]=A[s_p\tau]&\;\;&\textrm{(Annealing Pause)}\;\; t\in[\tau,\tau+\rho]\nonumber\\
A^F[t]=A[(1-s_p)(t-\rho)-(1-2s_p)\tau]&\;\;&\textrm{(Forward Annealing)}\;\; 
t\in[\tau+\rho,2\tau+\rho]\nonumber
\end{eqnarray}
where $\rho$ is the duration of the pause and $s_p\in[0,1]$ indicates the
location of the forward schedule where the pause is implemented. The total
duration of the selected reverse anneal protocol is $2\tau+\rho$ as opposed to
$\tau$ for the forward anneal.

While the theory of reverse annealing is just starting to be investigated~\cite{marshall2018pausing,ohkuwa2018reverse,kechedzhi2018efficient},
the physics rationale of reverse annealing is to be found in the oversimplified idea that, 
if the system is initialized in a state $\mathcal{S}$ corresponding to a local minimum of the 
objective function (\ref{eq:IsingObjFun}), the interplay of quantum and thermal fluctuations might 
help the state \emph{tunnel} out of the energy trap during the reverse annealing, while the 
annealing pause (and to some extent the final forward annealing) allows the system to thermalize 
and relax in the neighborhood of the newfound minimum. See Figure~{\ref{fig:schedule}-c} for a 
pictorial view of the concept.

The quality of the initial state $\mathcal{S}$ is likely to influence dramatically the 
reverse annealing process. For our experiments we use a classical greedy algorithm to
set $\mathcal{S}$, as described in the next section.\\

\section{Experimental Results} \label{sec:main}

We aim to solve representative portfolio instances at the limit of programmability of the 
most advanced DW2000Q device, and to benchmark the results by means of a competitive heuristics 
running on classical computers. Note that due to the embedding overhead  on DW2000Q we can embed 
a maximum of 64 logical binary variables on a fully-connected graph, which means that the largest 
search space for our benchmarks is around $60!/(30!)^2\simeq$ hundred thousand trillion if $M=N/2$. 
Our instance set consists of 30 randomly generated instances for each of the following number of 
assets $N=$\{24, 30, 36, 42, 48, 54, 60\}. In Table \ref{tab:GreedySearchAndParams}
we show the expected number of assets in the optimal portfolio of our unconstrained 
problem sets, computed by using all methods in this study, showing that the search space is non-trivial for all tested cases. 
Since exhaustive search is out of question for the largest problems, we provide two 
algorithms below that will be used to have a reference benchmark of the DW2000Q results.

A common metric to benchmark performance of non-deterministic iterative heuristics
against quantum annealing is Time-to-Solution (TTS). The TTS is defined as the 
expected number of independent runs of the method in order to find the ground state
with probability (confidence level) $\alpha$:
\begin{equation} \label{TTS}
\textrm{TTS} = t_{\textrm{run}} \frac{\ln(1 - \alpha)}{\ln(1 - p)} \; ,
\end{equation}
where $t_{\textrm{run}}$ is the running time elapsed for a single run (either $\tau$ for 
forward or $2\tau+\rho$ for reverse), and $p$ is the probability of finding the optimum of 
the objective function in that single shot~\cite{ronnow2014defining, mandra2018deceptive}.

\paragraph{Greedy Search}
The simplest reference benchmark to consider is the portfolio obtained with 
the best $M$ assets, neglecting correlations ($b_{ij}$=0)\footnote{
We note that optimal portfolios constructed through minimization of objective
function~(\ref{eq:QUBO}) and the number of asset constraints with QUBO
coefficients given by Table~\ref{tab:tableBuckets} have typically  better
Sharpe ratios than alternative portfolios constructed from the individually
best assets where the $a_i$ coefficients have not been coarse-grained in
buckets.}. 
The second simplest approach is to consider an iterative greedy search 
where correlations are taken into account progressively starting from the
selection/exclusion of the most desirable/undesirable assets. This is
implemented for instance by the routine shown in the pseudocode 
(Algorithm~\ref{alg:GSIH}) below\footnote{This algorithm is inspired by 
the routine provided by D-Wave Systems to decode the binary value of a set 
of qubit measurements that are originally associated to a single logical
variable $s_i$ (i.e., the $N_c$ spins ferromagnetically coupled during 
embedding $-$ see Eq.(\ref{eq:IsingEmbeddedObjFunJF}) and Ref.\cite{king2014algorithm}).}.

Note that this greedy search is the procedure we use to set the initial state
$\mathcal{S}$ for reverse annealing. Table~\ref{tab:GreedySearchAndParams} displays
results obtained by applying the greedy search algorithm to the unconstrained
problem Eq.~(\ref{eq:QUBO}) for the instance set, as a reference point,
highlighting the percentage of instances solved with the greedy search heuristic 
as a function of problem size.

\begin{table}[h!]
\centering
\footnotesize
\begin{tabular}{c | c c | c | c c c}
\toprule
{\bf Problem} & \multicolumn{2}{c|}{\bf \% of instances solved:} & {\bf Number of assets} & 
\multicolumn{3}{c}{\bf Parameters search space} \\
{\bf size}    & {\bf Greedy} & {\bf Forward}    & {\bf in the optimal} & \\
{\bf $N$}     & {\bf Search} & {\bf Annealing}  & {\bf portfolio}      & 
{\bf $J_F$}   & {\bf $s_p$}  & {\bf $\rho$ (\textmu$\textrm{s}$)} \\
\midrule
  24    &   80\%    &  100\%    & 10 ($-3,+$4)   &  3-7(0.5) & $-$ & $-$ \\
  30    &   90\%    &  100\%    & 12 ($-5,+$7)   &  3-7(0.5) & $-$ & $-$ \\
  36    &   93\%    &  97\%     & 13 ($-7,+$7)   &  3-7(0.5) & $-$ & $-$ \\
  42    &   70\%    &  90\%     & 16 ($-7,+6$)   &  5-8(0.5) & 0.32-0.5(0.02) & 1-15(7) \\
  48    &   47\%    &  87\%     & 17 ($-6,+5$)   &  5-8(0.5) & 0.32-0.5(0.02) & 1-15(7) \\
  54    &   60\%    &  60\%     & 19 ($-7,+12$)  &  5-8(0.5) & 0.32-0.5(0.02) & 1-15(7) \\
  60    &   57\%    &  30\%     & 23 ($-13,+15$) &  6-9(0.5) & 0.32-0.5(0.02) & 1-15(7) \\
\bottomrule
\end{tabular}
\caption{\small{Benchmark Instance Set characterization. For the number of assets in the optimal portfolio we report the median and the maximum and minimum in parenthesis, over 30 instances. The parameters search space for optimization of quantum annealing is reported in the last three columns, in the format min-max(step). } }
\label{tab:GreedySearchAndParams}
\end{table}

{\centering
\begin{minipage}{1.0\linewidth}
\begin{algorithm}[H]
\caption{Greedy Search Heuristic}\label{alg:GSIH}
\small{
\begin{algorithmic}[1]
\STATE \# Initialize energy tuples with the local magnetic fields, $h$:
\FOR{$i$ {\bf from} 0 {\bf to} {\it number of qubits} $-$ 1}
\STATE $energy[i]$ = $\{-|h(i)|, h(i), i\}$
\ENDFOR
\STATE \# Reorder energy tuples into a heap according to absolute magnitude:
\STATE $Energies$ = heap($energy$)
\STATE \# Initialize solutions:
\WHILE{$Energies$}
\STATE \# Energy tuple elements (largest magnitude energy tuple first):
\STATE $\{x, e, i\}$ = heappop($Energies$)
\IF{$e>0$} 
\STATE $Solution[i] = -1$
\ELSE 
\STATE $Solution[i] = +1$
\ENDIF
\STATE \# Update the rest of the heap:
\FOR{$z$ {\bf in} $Energies$}
\STATE $n = z[2]$ \# Qubit number
\STATE \# Energy update with the coupling strength, $J$:
\STATE $z[1] = z[1] + Solution[i] * (J(i,n) + J(n,i))$
\STATE $z[0] = -|z[1]|$
\ENDFOR
\ENDWHILE
\end{algorithmic}
}
\end{algorithm}
\end{minipage}
}
\\

\paragraph{Genetic Algorithm} We choose Genetic Algorithm (GA) as a
classical benchmark heuristics. Genetic Algorithms are adaptive methods of
searching a solution space by applying operators modelled after the natural
genetic inheritance and simulating the Darwinian struggle for survival. 
There is a rich history of GA applications to solving portfolio optimization
problems \cite{lin2005genetic, oh2005using}, including recent researches 
\cite{Kondratyev2017risk, kshatriya2018genetic} 
that explored a range of new portfolio optimization use cases.

A GA performs a multi-directional search by maintaining a population of
proposed solutions (called chromosomes) for a given problem. Each solution is
represented in a fixed alphabet with an established meaning (genes). The
population undergoes a simulated evolution with relatively good solutions
producing offsprings, which subsequently replace the worse ones. The estimate 
of the quality of a solution is based on a fitness function, which plays 
role of an environment. The simulation cycle is performed in three basic steps.
During the selection step a new population is formed by stochastic sampling
(with replacement). Then, some of the members of the newly selected 
populations recombine. Finally, all new individuals are re-evaluated. The 
mating process (recombination) is based on the application of two operators:
mutation and crossover. Mutation introduces random variability into the
population, and crossover exchanges random pieces of two chromosomes in the 
hope of propagating partial solutions.

In the case of portfolio optimization problem the solution (chromosome) 
is a vector ${\bf q} = (q_1, \ldots, q_N)$ consisting of $N$ elements (genes)
that can take binary values $q_i \in \{0, 1\}$. Our task is to find the
combination of genes that minimizes the objective (fitness) function $O({\bf
q})$. Due to the relatively short string of genes we do not use the crossover
recombination mechanism as it provides very little value in improving the
algorithm convergence (see Algorithm~\ref{alg:GAU} for details).\\

{\centering
\begin{minipage}{1.0\linewidth}
\begin{algorithm}[H]
\caption{Genetic Algorithm $-$ unconstrained portfolio optimization}
\label{alg:GAU}
\begin{algorithmic}[1]
\STATE Generation of $L$ initial solutions by populating the chromosomes 
       through the random draw from the pool of possible gene values \{0, 1\}.
\STATE Evaluation of the objective (fitness) function for each solution.
\STATE Ranking of solutions from 'best' to 'worst' according to the objective
       function evaluation results.
\FOR{$i$ {\bf from} 0 {\bf to} {\it number of iterations} $-$ 1}
\STATE Selection of $K$ best solutions from the previous generation and
       production of $L$ new solutions by randomly changing the values of 
       one or more genes. With $L = mK$ every one of the 'best' solutions 
       will be used to produce $m$ new solutions.
\STATE Evaluation of the objective (fitness) function for each solution.
\STATE Ranking of solutions from 'best' to 'worst' according to the objective
       function evaluation results.
\ENDFOR
\end{algorithmic}
\end{algorithm}
\end{minipage}
}
\\

The best values of parameters $L$ and $K$ depend on the problem size and specific QUBO 
coefficient mapping scheme and can be found through the trial and error process.
The objective here is to achieve the target convergence with the smallest number of 
objective function calls. The computational time of a single objective function call 
on the test machine (Intel(R) Xenon(R) CPU E5-1620 v4 processor run at 20\% CPU utilization) 
has been measured to be proportional to $N^2$ up to a cost of about 30 \textmu$\textrm{s}$ 
for $N=60$.

Note that, as with all heuristics, the GA approach does not
guarantee optimality. Hence the scaling measured is not necessarily
representative of the complexity of the problem $-$ but rather a measure
of the ability of the algorithmic approach to find the best solution.

\paragraph{Optimization of Forward and Reverse Quantum Annealing}
As discussed in Section \ref{sec:annealer}, our runs on DW2000Q are fully
specified by: an embedded Ising model (\ref{eq:IsingEmbeddedObjFunJF}) where the
free embedding parameter $J_F$ has been set, an annealing time $\tau$, and the
pause time $\rho$ and location $s_p$ (for reverse annealing). 
Moreover, in order to obtain meaningful results on the values of the logical
variables, a \emph{majority voting decoding} procedure has been applied to the
returned measurements of the physical qubits of the embedded Ising Eq.~(\ref{eq:IsingEmbeddedObjFunJF}-\ref{eq:IsingEmbeddedObjFun}) for each 1D chain,
as customary~\cite{venturelli2015spinglass}. Finally, runs are separated in batches
of 1500 anneals each with a different spin-reversal transormation (or "gauge") in order
to average out systematic errors during the anneals~\cite{king2014algorithm}.

\begin{figure}[H] 
\centering 
\begin{tikzpicture} 
\tikzstyle{every node}=[font=\footnotesize]
\begin{axis}[%
ymin = 2.5, ymax = 10,
xmin = 22, xmax = 62,
xlabel={Problem size},
ylabel={Optimal $J_F$},
legend style={draw=none},
legend style={cells={anchor=west}, legend pos=north west, fill=none},
width=0.5\textwidth, height=0.5\textwidth,
scatter/classes={%
a={mark=square*, blue1, mark size=2pt},
b={mark=square*, green1, mark size=4pt},
c={mark=square*, yellow1, mark size=4pt},
d={mark=square*, orange1, mark size=4pt},
e={mark=square*, red1, mark size=4pt},
f={mark=square*, gray1, mark size=4pt},
am={mark=square*, blue2, mark size=2pt},
bm={mark=square*, green2, mark size=4pt},
cm={mark=square*, yellow2, mark size=4pt},
dm={mark=square*, orange2, mark size=4pt},
em={mark=square*, red2, mark size=4pt},
fm={mark=square*, gray2, mark size=4pt},
al={mark=square*, blue3, mark size=2pt},
bl={mark=square*, green3, mark size=4pt},
cl={mark=square*, yellow3, mark size=4pt},
dl={mark=square*, orange3, mark size=4pt},
el={mark=square*, red3, mark size=4pt},
fl={mark=square*, gray3, mark size=4pt},
g={mark=square*, black, mark size=2pt}
}]
\addplot[
mark=o, mark size=4pt, green, thick
]
coordinates{(42, 6.0)(48, 7.0)(54, 7.0)(60, 8.0)};                      
\addplot[
mark=o, mark size=3.7pt, violet, thick
]
coordinates{(42, 7.0)(48, 6.5)(54, 7.5)(60, 7.0)};      
\addplot[
mark=o, mark size=4.8pt, red, thick
]
coordinates{(42, 6.0)(48, 6.5)(54, 7.5)(60, 8.5)};    
\addplot[        
mark=square*, mark size=2pt, black, line width=1.2pt
]  
coordinates {
(24, 3.5)(30, 4.5)(36, 5.5)(42, 6)(48, 7)(54, 7)
};
\addplot[scatter,only marks,%
scatter src=explicit symbolic]%
table[meta=label] {
x y label
24 3.0 a
24 3.5 g
24 4.0 a
24 4.5 a
24 5.0 am
24 6.0 al
30 3.0 a
30 4.0 a
30 4.5 g
30 5.0 a
30 5.5 a
30 6.0 am
30 6.5 al
36 3.5 am
36 4.5 a
36 5.0 al
36 5.5 g
36 6.0 a
36 6.5 a
36 7.0 a
42 5.0 am
42 5.5 a
42 6.0 g
42 6.5 am
42 7.0 am
42 7.5 al
42 8.0 a
48 5.5 al
48 6.0 al
48 6.5 a
48 7.0 g
48 7.5 am
48 8.0 am
54 5.5 a
54 6.0 a
54 6.5 al
54 7.0 g
54 7.5 a
54 8.0 a
};
\legend{{Reverse QA: $\rho$ = 1 \textmu$\textrm{s}$},
        {Reverse QA: $\rho$ = 8 \textmu$\textrm{s}$},
        {Reverse QA: $\rho$ = 15 \textmu$\textrm{s}$},
        {Forward QA}
       } 
\end{axis}
\end{tikzpicture}%
~%
\begin{tikzpicture} 
\tikzstyle{every node}=[font=\footnotesize]
\begin{axis}[
ymin = 0.32, ymax = 0.5,
xlabel={Pause time, $\rho$ (in \textmu$\textrm{s}$)}, ylabel={Optimal $s_p$},
symbolic x coords={1, 8, 15},
xtick=data,
xticklabel style={text width=0.5cm},
legend style={draw=none},
width=0.4\textwidth, height=0.5\textwidth,
legend style={cells={anchor=west}, legend pos=south east, fill=none}
]
\addplot[
mark=*,
mark size=2pt,
blue,
thick,
error bars/.cd, y dir=both, y explicit,
]
plot coordinates{ 
				 (1,  0.387) -= (1,  0.054) += (1,  0.054)
                 (8,  0.442) -= (8,  0.025) += (8,  0.025)
				 (15, 0.451) -= (15, 0.025) += (15, 0.025)
                }; 
\addplot[
mark=square*,
mark size=2pt,
red,
thick,
error bars/.cd, y dir=both, y explicit,
]
plot coordinates{ 
				 (1,  0.379) -= (1,  0.021) += (1,  0.021)
                 (8,  0.433) -= (8,  0.014) += (8,  0.014)
				 (15, 0.440) -= (15, 0.011) += (15, 0.011)
                };      
\addplot[
mark=diamond*,
mark size=3pt,
green,
thick,
error bars/.cd, y dir=both, y explicit,
]
plot coordinates{ 
				 (1,  0.384) -= (1,  0.026) += (1,  0.026)
                 (8,  0.436) -= (8,  0.031) += (8,  0.031)
				 (15, 0.445) -= (15, 0.027) += (15, 0.027)
                };           
\addplot[
mark=triangle*,
mark size=3pt,
orange,
thick,
error bars/.cd, y dir=both, y explicit,
]
plot coordinates{ 
				 (1,  0.366) -= (1,  0.030) += (1,  0.030)
                 (8,  0.444) -= (8,  0.044) += (8,  0.044)
				 (15, 0.447) -= (15, 0.040) += (15, 0.040)
                };                
\legend{{$N=42$},
        {$N=48$},
        {$N=54$},
        {$N=60$}
       } 
	\end{axis} 
\end{tikzpicture}
\caption{\small{Left: Median and distribution of optimal $J_F$ as a function of
problem size (See Table~\ref{tab:GreedySearchAndParams} for ranges of search). Darker colors indicate higher probability (forward quantum
annealing results). Right: Mean optimal $s_p$ as a function of pause time, $\rho$.
Error bars show 1 standard deviation confidence bands. 
Annealing time $\tau$ was fixed at 1~\textmu$\textrm{s}$.
}}
\label{fig:Optimal_JF_sp}
\end{figure}
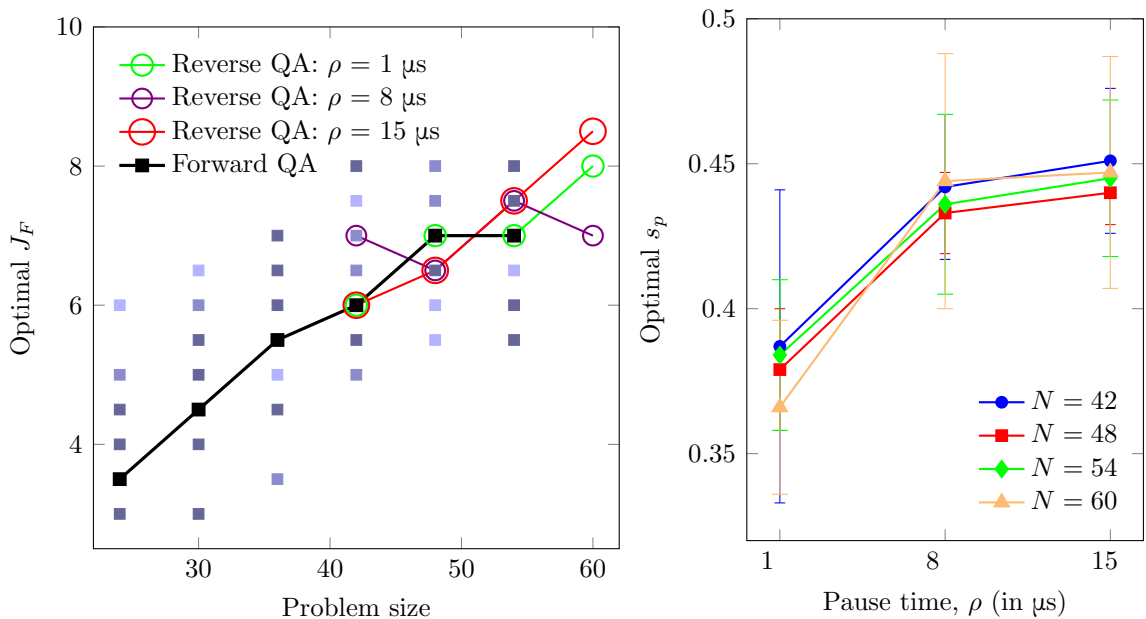

In order to determine the best parameters, we bruteforce over a fixed set of values 
(see Table \ref{tab:GreedySearchAndParams}), and we measure the TTS for all those
parameters, singling-out the lowest found value, instance by instance 
(see Appendix \ref{sec:ra}). This simple procedure 
represents a statistical pre-characterization of the instance ensemble sensitivity 
to the parameters, and it is common for quantum annealing benchmarking initial
studies of a problem~\cite{venturelli2015quantum,venturelli2015spinglass,hamerly2018scaling}.
Due to the post-selection of the best embedding and pause location
parameters, the reported results then will represent a “best case scenario”,
however previous empirical research show that once the distribution of typical
optimal parameters has been estimated, it is possible to set up a lightweight
on-the-fly procedure to set the parameters dynamically while performing the
runs~\cite{perdomo2015performance}.
The results of the parameter setting optimization are shown in Figure
\ref{fig:Optimal_JF_sp} for forward and reverse annealing.
It should be immediately noted that we don’t optimize over $\tau$ since our 
experiments (as well as prior research on similar instances~\cite{hamerly2018scaling}) 
indicates that the best TTS is obtained for the fastest annealing time, 
$\tau$ = 1~\textmu$\textrm{s}$~\footnote{for the largest problems studied, 
according to~\cite{hamerly2018scaling} there might be an advantage in varying $\tau$, 
but this is usually a small prefactor. See Also appendix \ref{sec:ra} for results on a limited set of instances.}.
What is shown is that the median best choice for $|J_F|$ is increasing 
with the problem size (as expected~\cite{venturelli2015spinglass}) and the mean best choice for
$s_p$ increases with the pause period. With regard to reverse annealing, we 
should note that we don’t pre-characterize all 30 instances, but only those 
that have not been solved by the greedy heuristics (Algorithm~\ref{alg:GSIH}) 
that is used to initialize the reverse annealing procedure. For those instances 
we are facing a three-parameter optimization ($|J_F|$, $\rho$, $s_p$).

What is clear from the results on reverse annealing is that (i) a good guess for 
the optimal $J_F$ parameter is accurately predicted by 
the value obtained for forward annealing, and (ii) the optimal pause location 
is consistently later in the anneal for larger $\rho$.
The first observation considerably simplifies the parameter setting 
strategy when considering operational scenarios with a new unknown
instance, while the second observation is consistent with recent results 
and interpretation of the annealing pause tested on native Chimera
problems~\cite{marshall2018pausing}.

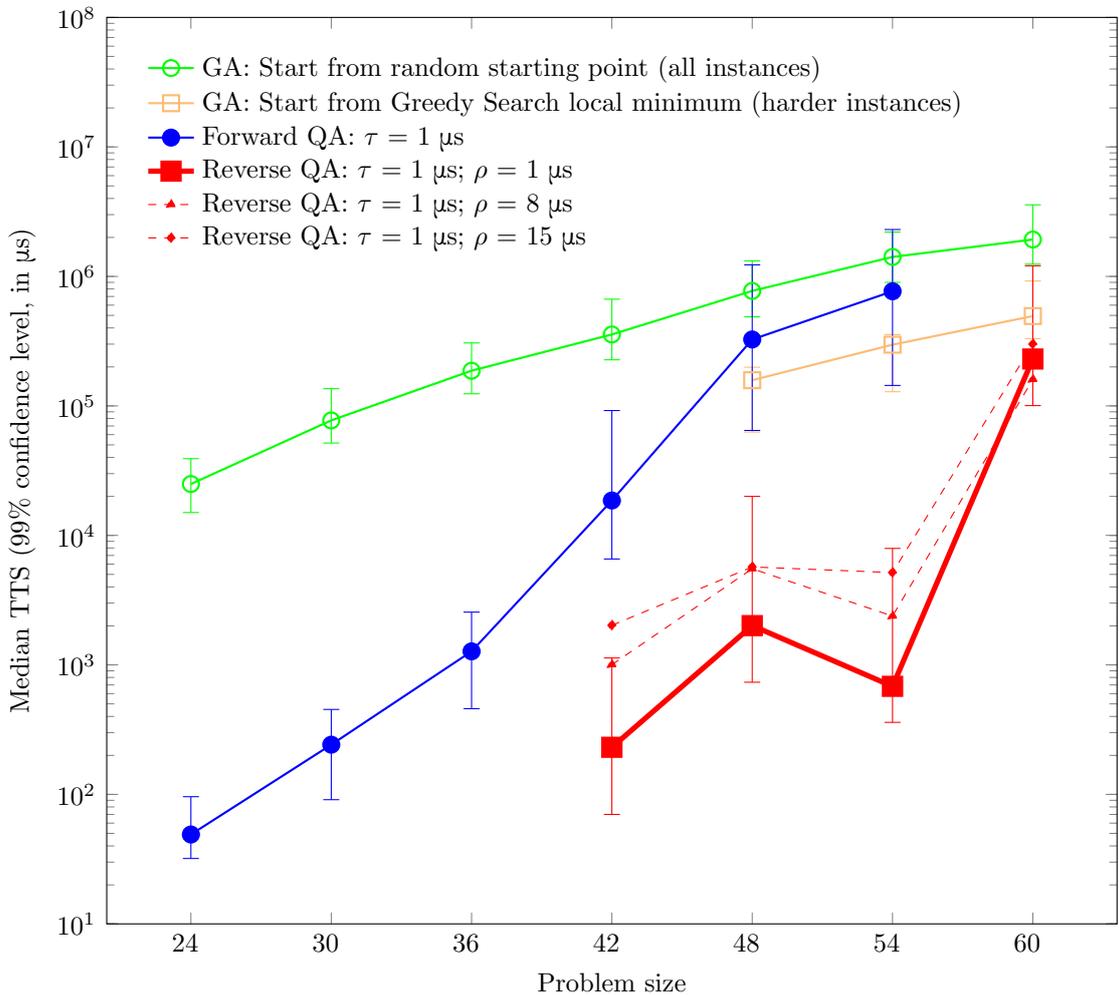
\begin{figure}[H] 
\centering 
\begin{tikzpicture} 
	\tikzstyle{every node}=[font=\footnotesize]
	\begin{semilogyaxis}[
     ymin = 10, ymax = 100000000,
     xlabel={Problem size}, ylabel={Median TTS (99\% confidence level, in \textmu$\textrm{s}$}),
     symbolic x coords={24, 30, 36, 42, 48, 54, 60},
     xtick=data,
     xticklabel style={text width=0.5cm},
     legend style={draw=none},
     width=0.875\textwidth, height=0.8\textwidth,
     legend style={cells={anchor=west}, legend pos=north west, fill=none}
		]
     \addplot[
        mark=o,
        mark size=3pt,
        green,
        thick,
        error bars/.cd, y dir=both, y explicit,
        ]
        plot coordinates { 
						  (24,    24938) -= (24,    9920) += (24,   14230)
                          (30,    77285) -= (30,   25590) += (30,   58479)
						  (36,   186990) -= (36,   62937) += (36,  119506)
						  (42,   356183) -= (42,  129194) += (42,  310955)
						  (48,   772771) -= (48,  285426) += (48,  545803)
						  (54,  1413780) -= (54,  513774) += (54,  783389)
						  (60,  1926585) -= (60,  683965) += (60, 1647030)
                         }; 
     \addplot[
        mark=square,
        mark size=3pt,
        orange,
        thick,
        error bars/.cd, y dir=both, y explicit,
        ]
        plot coordinates { 
						  (48,   157922) -= (48,  95097) += (48,   41075)
						  (54,   296979) -= (54, 168291) += (54,   58750)
						  (60,   494104) -= (60, 163040) += (60,  427373)
                         };                          
     \addplot[
        mark=*,
        mark size=3pt,
        blue,
        thick,
        error bars/.cd, y dir=both, y explicit,
        ]
        plot coordinates { 
						  (24,       49) -= (24,	  17) += (24,      47)
                          (30,      242) -= (30,	 151) += (30,      211)
						  (36,     1272) -= (36,	 813) += (36,     1284)
						  (42,    18603) -= (42,	12027) += (42,    73497)
						  (48,   325902) -= (48,  261270) += (48,  902141)
						  (54,   767526) -= (54,  623617) += (54, 1535060)
                         }; 
     \addplot[
        mark=square*,
        mark size=3pt,
        red,
        thick,
        line width=2pt,
        error bars/.cd, y dir=both, y explicit,
        ]
        plot coordinates { 
						  (42,      231) -= (42,	 161) += (42,      900)
						  (48,     2010) -= (48,    1275) += (48,    18010)
						  (54,      683) -= (54,     323) += (54,    7254)
						  (60,   230256) -= (60,  129724) += (60,  976974)
                         }; 
     \addplot[
        mark=triangle*,
        red,
        dashed,
        error bars/.cd, y dir=both, y explicit,
        ]
        plot coordinates { 
						  (42,     1004) 
						  (48,     5527) 
						  (54,     2383) 
						  (60,   160161) 
                         }; 
     \addplot[
        mark=diamond*,
        red,
        dashed,
        error bars/.cd, y dir=both, y explicit,
        ]
        plot coordinates { 
						  (42,     2021) 
						  (48,     5709) 
						  (54,     5163) 
						  (60,   300303) 
                         }; 
		\legend{{GA: Start from random starting point (all instances)},
		        {GA: Start from Greedy Search local minimum (harder  instances)},
		        {Forward QA: $\tau$ = 1 \textmu$\textrm{s}$},
		        {Reverse QA: $\tau$ = 1 \textmu$\textrm{s}$; $\rho$ = 1 \textmu$\textrm{s}$},
		        {Reverse QA: $\tau$ = 1 \textmu$\textrm{s}$; $\rho$ = 8 \textmu$\textrm{s}$},
		        {Reverse QA: $\tau$ = 1 \textmu$\textrm{s}$; $\rho$ = 15 \textmu$\textrm{s}$}
		       } 
	\end{semilogyaxis} 
\end{tikzpicture}
\caption{\small{Time-to-solution (99\% confidence level): GA, Forward and Reverse Quantum 
Annealing. Curves follow the median and error bars indicate 30\% and 70\% percentile over the 30 
instances. The TTS curve for Forward QA is limited to $N=54$ since for $N=60$ the median is not 
well defined, as only 30\% of instances solve to the best known objective value. Dashed curves 
indicate median TTS obtained with higher $\rho$ (no error bars displayed for clarity). All TTS 
are measured not counting the time required to run the greedy descent that initializes the initial ansatz $\mathcal{S}$, nor the overhead times for operating the DW2000Q 
(see Discussion in Section~\ref{sec:conclusion}).}}
\label{fig:FQAvsRQA}
\end{figure}

\paragraph{Results}
Figure \ref{fig:FQAvsRQA} displays the TTS results for the GA algorithm, the forward 
QA solver and the reverse QA solver results. Of particular interest of this study are 
the probability of success obtained for 42-, 48-, 54- and 60-asset portfolios against the 
forward annealing results obtained for all portfolio sizes.
We report also on the comparison with GA - to have a baseline reference on how 
comparatively perfomant is the quantum solver against a classical CPU. GA can also be
initialized by the Greedy Search heuristics, and this also decreases the TTS required for GA to 
find the global minimum (orange curve in Figure~\ref{fig:FQAvsRQA}).

From the comparisons, the best solver appears to be reverse quantum annealing at 
minimum annealing time and pause time. In the median case, we observe one to three orders of 
magnitude speed-up when applying reverse quantum annealing with respect to quantum 
annealing\footnote{We believe that the non-monotonic behavior for $N=54$ is not of 
fundamental significance but it is due to the finite small size of our instance set for 
reverse annealing.}.

\section{Conclusion and Next Steps} \label{sec:conclusion}

In this paper we did investigate a combinatorial optimization problem in
Finance, identifying parameters and mappings that are related to real-world
portfolio optimization approaches and that are compatible with the precision
requirements of current D-Wave machines. The resulting QUBO has integer
coefficients and is very densely connected. Hence, we leveraged prior research 
on fully-connected graphs by employing the state-of-the-art methods to compile
the problem on the Chimera architecture.

While for general coefficients the quadratic problem is NP-hard, the
instances that we randomly generate are not necessarily hard. As a matter of
fact, using the simple greedy pre-processing approach (Algorithm \ref{alg:GSIH})
we can solve a large part of them as shown in Table~\ref{tab:GreedySearchAndParams}. 
We did not filter on difficulty since we don’t expect all portfolio optimization
problems of interests to be hard necessarily, and it is difficult to classify
the instances on their expected difficulty beforehand. However, as the number of
variables $N$ increases, while QUBO parameters are kept the same, we expect
greedy approaches and classical heuristics to fail on typical instances.

Consistently with the rest of the literature, all reported times for runs
on the DW2000Q have been calculated in terms of the annealing times $\tau$ (or
$\rho$), ignoring all other necessary setup and iteration times, since they
don’t scale with the problem size and they could be reduced in the future as
system integration of the chips improves. However, once one considers the
additional overhead times required to complete a full run on the D-Wave
machine\footnote{programming time, post-programming thermalization time, readout time; 
respectively 7.575 ms, 1 ms, 124.98 $\mu$s for the current experiments}, 
the advantage disappears, and the TTS results are on par with our
classical benchmark. It is important to note that the reported scaling of TTS with problem size has
not a fundamental character as the parameters that influence TTS ($J_F$, $\tau$, $\rho$, $s_p$) 
are optimized only within a limited range. In particular, the annealing time is expected to be
suboptimal for $N \leq 50$, leading to a measurement of a flatter slope for the median
TTS~\cite{albash2018demonstration}. While asymptotic scaling analysis can be extrapolated only
on future devices that support more variables and parameter values, we believe there is value
in reporting on the complexity according to the current operational scenarios. 

We note that improvements can be made
over our analysis and approach and engineering advances are being
planned in the next few years on quantum annealers. For instance, research
papers on reverse annealing and paused annealing are just beginning to appear~\cite{marshall2018pausing,ottaviani2018low}, and it seems that for hard 
problems it might be beneficial to implement a longer pause, larger than 100 
microseconds, to gain a sizeable TTS advantage. Moreover, we do not expect 
difference in probability of success if we were to decrease the forward/reverse 
anneal time $\tau$ beyond 1 \textmu$\textrm{s}$, since the 
process is dominated by thermalization during the pause.
Based on these considerations, it is reasonable to believe that more than an order 
of magnitude of performance can be gained by further tuning the quench time $\tau$ 
and the pause time $\rho$. 

A second very promising development that could also lead to 
order of magnitudes improvements is the transition of the D-Wave Chimera architecture 
to the Pegasus architecture. According to forecasts, the next chip will be able to 
embed almost 400 logical variables, reducing embedding overhead of at least a factor 
of 3. According to recent research~\cite{hamerly2018scaling} the embedding overhead 
is responsible for a large part of the performance, hence we expect conservatively
to gain at least an order of magnitude from this improvement. 
Another possible performance enhancement could be coming from
inhomogeneous driving~\cite{inhomDrivingQCWARE}.
As the number of 
physical qubits increases, we could also leverage error-suppression encodings
nested with embedding~\cite{vinci2016nested} that while reducing the total number 
of logical variables they provably improve the probability of success.
Additional knobs available that could improve upon the current TTS are: 
the use of multiple pauses and the additional precision on the J couplings offered
by the \emph{extended $\textrm{J-range}$} feature.

Bearing all these considerations in
mind, while it is not clear if quantum annealing is going to be the most
compelling solver for portfolio optimization, our results indicate that as
technology and theory progresses it could represent a viable choice. Our conclusion is indeed that these first results on the use of quantum annealing in 
reverse mode look very promising, especially in light of the large room for improvement that 
is possible to achieve by future hardware design.

From a Finance perspective, another direction of future research is to explore 
alternative portfolio optimization approaches. For example, we can mention here work 
done by Sortino and van der Meer \cite{sortino1991downside} on skewed return distributions. 
Sortino ratio that normalizes excess return by the downside deviation often provides more
valuable information about relative portfolio performances than widely used Sharpe ratio, 
where excess return is normalized by the standard deviation of both positive and negative 
portfolio returns.

\vskip2cm

\noindent
{\bf \large{Acknowledgements}}\\

\noindent
The collaboration between USRA and Standard Chartered Bank has been supported 
by the USRA Cycle 3 Program that allowed the use of the D-Wave Quantum Annealer 
2000Q\textsuperscript{TM}, and by funding provided by NSF award No. 1648832 
obtained in collaboration with QC-Ware. We acknowledge QC-Ware and specifically 
thank Eric Berger, for facilitating the collaboration and contributing to the runs 
on the D-Wave machine. D.V. acknowledges general support from NASA Ames Research Center and 
useful discussions with QuAIL research team. A.K. would like to thank David Bell and 
USRA for the opportunity to conduct research on the quantum annealer at QuAIL.

\vskip2cm

\noindent
{\bf \large{Disclaimer and Declaration of Interest}}\\

\noindent
The authors report no conflict of interest. The authors alone are responsible for the
content and writing of the paper. The opinions expressed are those of the authors and
do not necessarily reflect the views and policies of Standard Chartered Bank or the 
Universities Space Research Association. All figures are based on own calculations.\\

\noindent
This paper is for information and discussion purposes only and does not constitute
either an offer to sell or the solicitation of the offer to buy any security or any
financial instrument or enter into any transaction or recommendation to acquire or 
dispose of any investment.

\newpage
  
\bibliographystyle{unsrt}  

\appendix

\section{Geometric Brownian Motion} \label{sec:gbm}

A Geometric Brownian Motion (GBM) is a stochastic process $S(t)$ that satisfies the following stochastic 
differential equation (SDE):
\begin{displaymath}
dS(t) = \mu S(t) dt + \sigma S(t) dB(t) \; ,
\end{displaymath}
where $t$ is continuous time and $B(t)$ is a Brownian motion. GBM is widely used to model asset prices.
If a unit of time is 1 year, then $\sigma$ is interpreted as an annualized volatility (standard deviation)
of asset's log-returns, which are assumed to be normally distributed. The drift coefficient $\mu$ controls
deterministic component of the asset price process. 

Integrating the process we obtain:
\begin{displaymath}
S(t) = S(0) \exp \left( \left(\mu - \frac{1}{2}\sigma^2 \right)t + \sigma B(t)\right) \; .
\end{displaymath}

Although GBM SDE can be used directly to simulate an asset process, it is better to use its solution to
ensure that simulated asset prices do not turn negative $-$ this may be the case for large enough time step. 
In our portfolio optimization example $\Delta t$ = 1 month and we use the following discretization scheme for 
a single asset price process:
\begin{displaymath}
S(t_{n}) = S(t_{n-1}) \exp \left( \left(\mu - \frac{1}{2}\sigma^2 \right)\Delta t + 
\sigma z_{n} \sqrt{\Delta t} \right) \; ,
\end{displaymath}
where $t_{n} = t_{n-1} + \Delta t$ and $z_{n}$ is a standard normal random variable. Asset prices from 
the $N$-asset portfolio are jointly simulated using the same scheme but correlated standard normal random 
variables $(z^{(1)}, \ldots, z^{(N)})$ are constructed via Cholesky decomposition of the 
correlation matrix $\rho$.\\

\section{Chimera Graph of DW2000Q and Embedding}
\begin{figure}[t]
\center
\includegraphics[width=0.75\columnwidth]{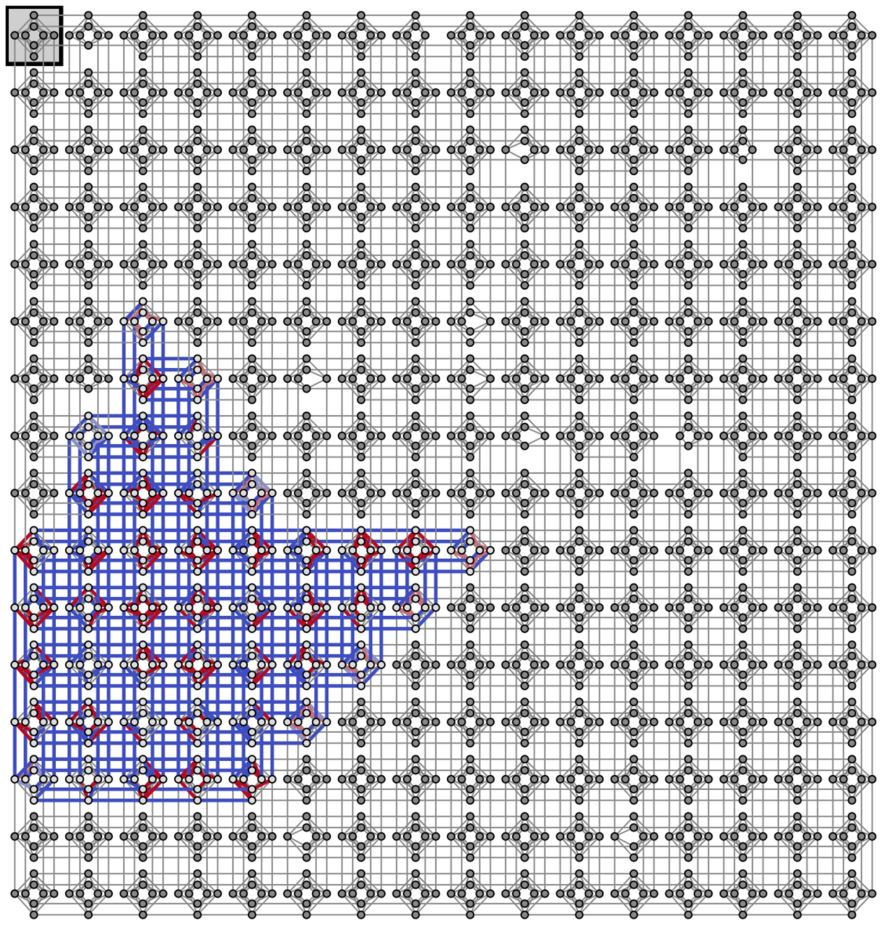}
\caption{\small{Chimera Chip of DW2000Q. Each gray dot represents an active qubit (missing dots are broken qubits), the black shaded square is representative of one unit cell. The embedding for an instance ($N$=42)
is highlighted: blue bonds are ferromagnetic couplings set to $J_F$, while red and pink bonds represent 
logical couplings ($J_{ij}$ in Eq.(\ref{eq:IsingEmbeddedObjFun})).}}
\label{fig:chimera}
\end{figure}

In Figure \ref{fig:chimera} we show the layout of the chip used for the experiments, 
belonging to the machine D-Wave 2000Q hosted at NASA Ames Research Center.

\label{appendixEmbedding}

\section{More details on parameter setting for Reverse Annealing} \label{sec:ra}

\begin{figure}
\centering 
\begin{tikzpicture} 
		\tikzstyle{every node}=[font=\footnotesize]
		\begin{semilogyaxis}[
     xlabel={Problem size}, 	
     ylabel={Median TTS (99\% confidence level), in \textmu$\textrm{s}$},
     symbolic x coords={24, 30, 36, 42, 48, 54},
     xtick=data,
     legend style={draw=none},
     width=0.7\textwidth, height=0.5\textwidth,
     legend style={cells={anchor=west}, legend pos=south east}
		]
     \addplot[
        mark=square*,
        blue,
        thick,
     ]  plot coordinates { 
         (24,     21)
         (30,    169)
         (36,    365)
         (42,   5179)
         (48,  38374)
         (54, 546051)
        }; 
     \addplot[
        mark=triangle*,
        magenta,
        thick,
     ]  plot coordinates { 
         (24,    127)
         (30,   1165)
         (36,   1720)
         (42,  10203)
         (48, 149281)
        }; 
				\legend{
              QA ($t_{\textrm{run}}$ = 1 \textmu$\textrm{s}$),
              QA ($t_{\textrm{run}}$ = 10 \textmu$\textrm{s}$)
             } 
		\end{semilogyaxis} 
\end{tikzpicture}
\caption{\small{Time-to-solution (99\% confidence level) for different $t_{\textrm{run}}$ (median over 10 instances).}}
\label{fig:TTS2}
\end{figure}
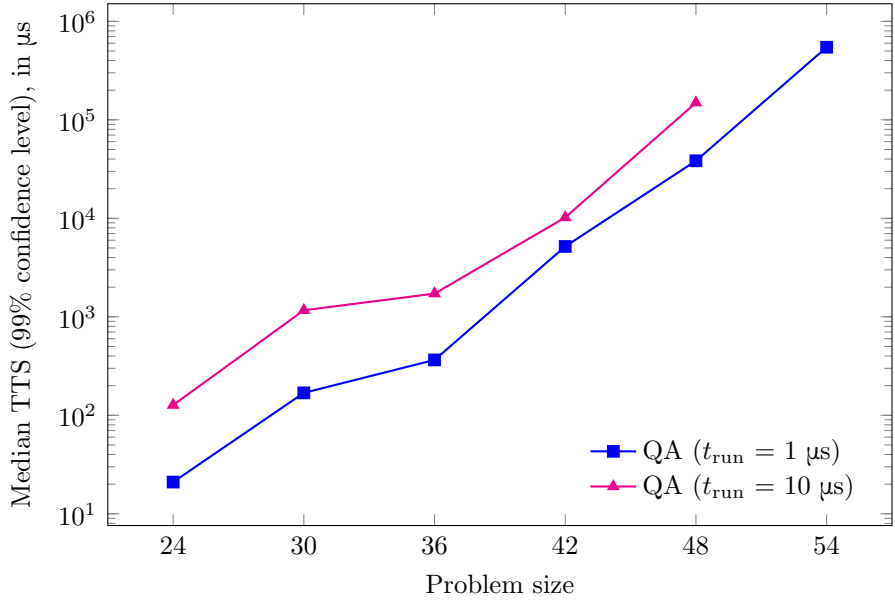

Figure \ref{fig:TTS2} 
displays median TTS results obtained for the mapping schemes provided by Table \ref{tab:tableBuckets} and 
annealing times 1 \textmu$\textrm{s}$ and 10 \textmu$\textrm{s}$, obtained for the first 10 instances of 
the benchmark ensamble on an independent set of runs with respect to the results presented in 
Figure~\ref{fig:FQAvsRQA}\footnote{The reported median TTS on these runs seems to be in general faster than the
results in the main paper. This could be due to finite statistics effect or to general drift in performance 
of the machine over time, since the runs relative to Fig.~\ref{fig:TTS2} were performed more than a month
earlier when the machine was under low utilization. The effective temperature of the machine can vary of 
few milliKelvins over time for uncontrollable factors, and this is known to affect the performance of 
quantum annealing~\cite{boixo2016computational}.}. It is clear that the choice of $\tau$ = 1
\textmu$\textrm{s}$ is the most advantageous.

\begin{figure} 
\centering 
\begin{tikzpicture} 
		\tikzstyle{every node}=[font=\footnotesize]
		\begin{semilogyaxis}[
     ymin = 100, ymax = 100000000,
     xmin = 0.3, xmax = 0.52,
     xlabel={$\rho$ = 1 \textmu$\textrm{s}$},
     ylabel={Median TTS (99\%) as a function of $s_p$, in \textmu$\textrm{s}$},
     width=0.35\textwidth, height=0.55\textwidth,
     legend style={cells={anchor=west}, legend pos=north west}
		]
     \addplot[        
        mark=diamond*,
        black,
     ]  
        coordinates {
       (0.32, 10006.6)(0.34, 2162.53)(0.36, 6094.96)(0.38, 1257.08)(0.40, 3045.17)(0.42, 5112.25)(0.44, 4181.91)(0.46, 5350.24)(0.48, 184202)(0.50, 184202)
     };
     \addplot[        
        mark=diamond*,
        blue,
     ]  
        coordinates {
       (0.32, 3320.43)(0.34, 9025.14)(0.36, 2055.87)(0.38, 961.845)(0.40, 996.512)(0.42, 921.981)(0.44, 3739.43)(0.46, 9393.7)(0.48, 460512)(0.50, 35419.8)
     };
     \addplot[        
        mark=diamond*,
        green,
     ]  
        coordinates {
        (0.32, 8445.25)(0.34, 282.744)(0.36, 2696.37)(0.38, 1080.23)(0.40, 868.405)(0.42, 750.951)(0.44, 5788.06)(0.46, 5477.74)(0.48, 102333)(0.50, 14615)
     };
     \addplot[        
        mark=diamond*,
        yellow,
     ]  
        coordinates {
       (0.32, 1497.89)(0.34, 2162.53)(0.36, 614.358)(0.38, 2642.04)(0.40, 1098.42)(0.42, 2665.06)(0.44, 4947.19)(0.46, 6972.92)(0.48, 65783.5)(0.50, 460512)
     };     
     \addplot[        
        mark=diamond*,
        orange,
     ]  
        coordinates {
       (0.32, 1984.67)(0.34, 2344.97)(0.36, 673.613)(0.38, 680.678)(0.40, 413.459)(0.42, 2928.62)(0.44, 7483.48)(0.46, 30696.5)(0.48, 32889.5)(0.50, 70844.2)
     };     
     \addplot[        
        mark=diamond*,
        red,
     ]  
        coordinates {
       (0.32, 5350.24)(0.34, 917.343)(0.36, 1416.74)(0.38, 1226.72)(0.40, 716.069)(0.42, 1716.95)(0.44, 2619.42)(0.46, 7026.19)(0.48, 23611.7)(0.50, 65783.5)
     };          
     \addplot[        
        mark=diamond*,
        brown,
     ]  
        coordinates {
       (0.32, 2010.78)(0.34, 700.618)(0.36, 1093.16)(0.38, 1156.84)(0.40, 658.953)(0.42, 911.839)(0.44, 5577.42)(0.46, 10344.1)(0.48, 15094.3)(0.50, 15875.3)
     };   
     
     \addplot[
        mark=o, mark size=4pt, black, line width=2pt
     ] coordinates {
       (0.34, 282.744)
     };

	\legend{
       $J_F = 5$,
       $J_F = 5.5$,
       $J_F = 6$,
       $J_F = 6.5$,
       $J_F = 7$,
       $J_F = 7.5$,
       $J_F = 8$
     } 
		\end{semilogyaxis} 
\end{tikzpicture}%
~%
\begin{tikzpicture} 
		\tikzstyle{every node}=[font=\footnotesize]
		\begin{semilogyaxis}[
     ymin = 100, ymax = 100000000,
     xmin = 0.3, xmax = 0.52,
     xlabel={$\rho$ = 8 \textmu$\textrm{s}$},
     legend style={draw=none},
     width=0.35\textwidth, height=0.55\textwidth,
     legend style={cells={anchor=west}, legend pos=north east, font=\scriptsize}
		]
     \addplot[        
        mark=diamond*,
        black,
     ]  
        coordinates {
       (0.40, 3684100)(0.42, 46894.8)(0.44, 4978.91)(0.46, 9059.74)(0.48, 2522.3)(0.50, 16824.2)
     };
     \addplot[        
        mark=diamond*,
        blue,
     ]  
        coordinates {
        (0.36, 263116)(0.38, 818660)(0.40, 22223.8)(0.42, 13865.5)(0.44, 5688.23)(0.46, 3394.92)(0.48, 3514)(0.50, 9243.05)
     };
     \addplot[        
        mark=diamond*,
        green,
     ]  
        coordinates {
       (0.36, 2456050)(0.38, 387767)(0.40, 54542.9)(0.42, 9519.88)(0.44, 12601.7)(0.46, 3468.39)(0.48, 15981.1)(0.50, 12844.7)
     };     
     \addplot[        
        mark=diamond*,
        yellow,
     ]  
        coordinates {
       (0.34, 2456050)(0.36, 1842030)(0.38, 31052.9)(0.40, 43562.4)(0.42, 4522.63)(0.44, 2762.49)(0.46, 2700.09)(0.48, 9243.05)(0.50, 7776.75)
     };     
     \addplot[        
        mark=diamond*,
        orange,
     ]  
        coordinates {
       (0.32, 1842030)(0.34, 1473620)(0.36, 230222)(0.38, 48438.6)(0.40, 8403.28)(0.42, 4525.46)(0.44, 1978.87)(0.46, 5774.02)(0.48, 9906.8)(0.50, 37556.4)
     };     
     \addplot[        
        mark=diamond*,
        red,
     ]  
        coordinates {
       (0.32, 818660)(0.34, 409312)(0.36, 71499.8)(0.38, 56209.5)(0.40, 9161.95)(0.42, 6244.65)(0.44, 7137.65)(0.46, 2614.4)(0.48, 4605.95)(0.50, 14758.9)
     };          
     \addplot[        
        mark=diamond*,
        brown,
     ]  
        coordinates {
       (0.32, 818660)(0.34, 526268)(0.36, 294694)(0.38, 116920)(0.40, 8042.35)(0.42, 4729.09)(0.44, 4123.56)(0.46, 3404.54)(0.48, 9048.52)(0.50, 28745.5)
     };         

     \addplot[
        mark=o, mark size=4pt, black, line width=2pt
     ] coordinates {
       (0.44, 1978.87)
     };

		\end{semilogyaxis} 
\end{tikzpicture}%
~%
\begin{tikzpicture} 
		\tikzstyle{every node}=[font=\footnotesize]
		\begin{semilogyaxis}[
     ymin = 100, ymax = 100000000,
     xmin = 0.3, xmax = 0.52,
     xlabel={$\rho$ = 15 \textmu$\textrm{s}$},
     legend style={draw=none},
     width=0.35\textwidth, height=0.55\textwidth,
     legend style={cells={anchor=west}, legend pos=north east, font=\scriptsize}
		]
     \addplot[        
        mark=diamond*,
        black,
     ]  
        coordinates {
       (0.42, 1726870)(0.44, 476328)(0.46, 47735.4)(0.48, 17485.5)(0.50, 8570.83)
     };
     \addplot[        
        mark=diamond*,
        blue,
     ]  
        coordinates {
       (0.40, 1973580)(0.42, 203100)(0.44, 19119)(0.46, 42571.3)(0.48, 9297.2)(0.50, 10878.1)
     };
     \addplot[        
        mark=diamond*,
        green,
     ]  
        coordinates {
        (0.36, 13815400)(0.38, 3453810)(0.40, 2302520)(0.42, 226415)(0.44, 4041.06)(0.46, 7259.89)(0.48, 4343.05)(0.50, 4045.96)
     };
     \addplot[        
        mark=diamond*,
        yellow,
     ]  
        coordinates {
       (0.32, 4605100)(0.34, 3453810)(0.36, 13815400)(0.38, 6907690)(0.40, 251122)(0.42, 83156.9)(0.44, 14322)(0.46, 14581.4)(0.48, 13843.7)(0.50, 32824.9)
     };     
     \addplot[        
        mark=diamond*,
        orange,
     ]  
        coordinates {
       (0.32, 13815400)(0.34, 418583)(0.36, 986753)(0.38, 170493)(0.40, 104594)(0.42, 13623.1)(0.44, 4604.3)(0.46, 13138.7)(0.48, 49987.1)(0.50, 26397.4)
     };     
     \addplot[        
        mark=diamond*,
        red,
     ]  
        coordinates {
       (0.32, 1973580)(0.34, 13815400)(0.36, 1534990)(0.38, 1973580)(0.40, 54972.8)(0.42, 18600.4)(0.44, 7152.61)(0.46, 13663.9)(0.48, 45079.6)(0.50, 15648.1)
     };          
     \addplot[        
        mark=diamond*,
        brown,
     ]  
        coordinates {
       (0.32, 6907690)(0.34, 3453810)(0.36, 627909)(0.38, 575577)(0.40, 54109.4)(0.42, 14322)(0.44, 10992)(0.46, 19499.6)(0.48, 35084.8)(0.50, 23033.7)
     };   
     \addplot[
        mark=o, mark size=4pt, black, line width=2pt
     ] coordinates {
       (0.44, 4041.06)
     };
     
		\end{semilogyaxis} 
\end{tikzpicture}
\caption{\small{Time-to-solution (99\% confidence level) as a function of annealing parameter $s_p$. 
Results for a single instance, $N = 42$, $\tau$ = 1~\textmu$\textrm{s}$.
Circles point out to the best found ($J_F$, $s_p$) for these three illustrative cases.}}
\label{fig:TTS_rho_JF}
\end{figure}
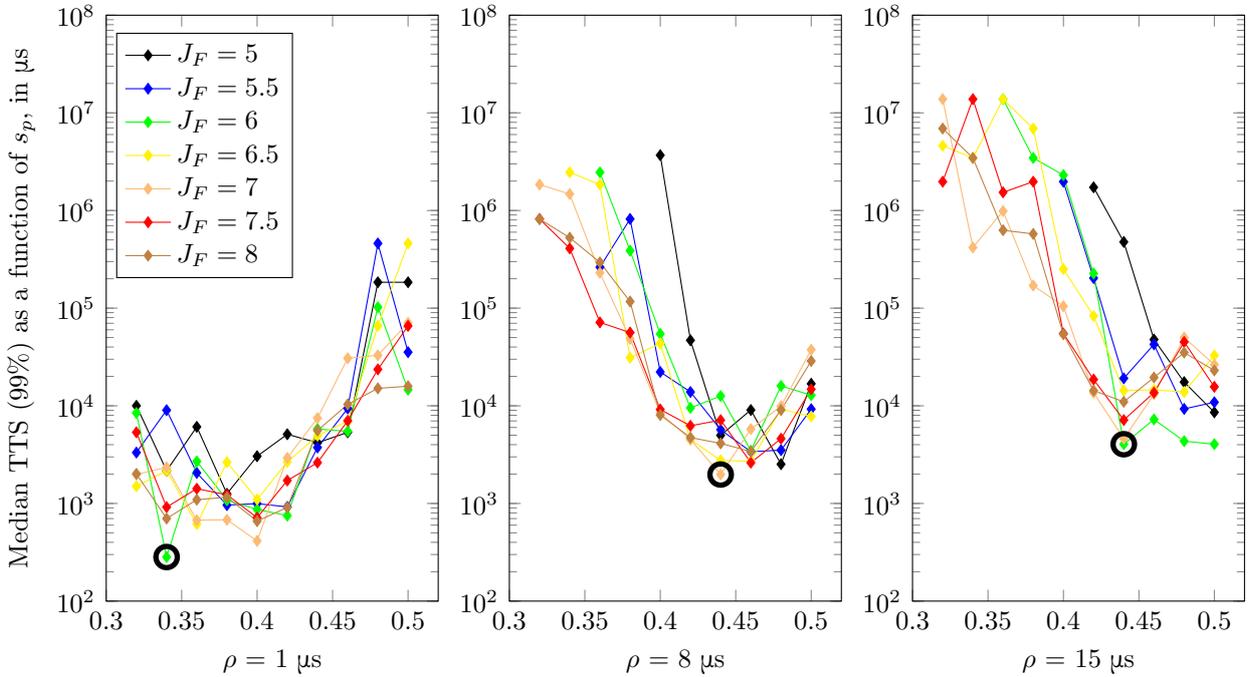

In Figure~\ref{fig:TTS_rho_JF} we show on an example how the optimal parameter setting is 
performed to generate data in Figure~\ref{fig:Optimal_JF_sp} and \ref{fig:FQAvsRQA}.
Scans are performed for different $J_F$ and $s_p$ and the best TTS is selected, 
instance by instance.

\end{document}